\documentclass[prl,preprint,preprintnumbers,showpacs,floatfix]{revtex4}
\usepackage{graphicx}
\usepackage{psfig}
\usepackage{dcolumn}
\usepackage{bm}
\usepackage{subfigure}
\usepackage{epsf}

\begin{document}

\title{How do drops evaporate?}

\author{N. Murisic}
\affiliation{Department of Mathematical Sciences,
Center for Applied  Mathematics and Statistics,
New Jersey Institute of Technology, Newark, NJ 07102}

\author{L. Kondic}
\affiliation{Department of Mathematical Sciences,
Center for Applied  Mathematics and Statistics,
New Jersey Institute of Technology, Newark, NJ 07102}

\date{\today}


\begin{abstract} 
We consider evaporation of pure liquid drops on
a thermally conductive substrate.  Two evaporative models are considered: 
one that concentrates on the liquid phase in determining evaporative flux, and
the other one that centers on the gas/vapor phase.  
A single governing equation for the evolution of drop thickness, including both models,
is developed.  Experiments are used to estimate relevant parameters.  
We show how the derived governing equation can be used to predict which 
evaporation model is appropriate under considered experimental conditions.
\end{abstract}

\maketitle

Evaporating thin films and drops are present in numerous natural situations
and applications of technical importance.  Coated liquid films, for example,
are often left to dry by evaporation.  The residual films, whose thickness may vary
from millimetric in the case of paints to nanometric for photoresist
films in semiconductor applications, are often desired in a uniform state.
However, various kinds of instabilities, many of them driven by evaporation-related
mechanisms, often occur.   Evaporative sessile drops are perhaps
even more interesting since nonuniform drop thickness and presence of contact lines
(separating liquid, gas, and solid phase) lead to new effects.  One of these is
a possibility of nonuniform evaporation along the gas-liquid interface, leading to temperature
gradients and related Marangoni effects.
These effects are crucial in a number of problems, including the so-called coffee-stain
phenomenon involving deposition of solid particles dissolved in the liquid close to a contact
line~\cite{DBDHNW1997}, and its' numerous applications, such as analysis
of DNA microarrays~\cite{blossey_natmat03}. 

Despite its apparent simplicity, the problem of an evaporating drop on a thermally 
conducting solid substrate involves a number of physical processes, including mass and energy 
transfer between the three phases, diffusion and/or convection of vapor in the gas phase, 
coupled to the complex physics in the vicinity of a contact line.
So-called `2-sided' models include the processes both in the liquid and in 
the gas phase, but lead to a mathematical formulation of significant complexity,
even when the solid phase and contact line issues are not considered~\cite{COLINET}.
Therefore, most of the researchers in the field have chosen to use 
various simplifications, allowing them to reduce the complete problem to a tractable mathematical
formulation.  As we discuss below, these simplifications are based on estimating the importance
of various physical processes, and lead eventually to models which concentrate on the 
processes in one of the phases (gas or liquid).  The estimates, however, involve quantities
which are either not known, or not known precisely enough.  We will show
that various assumptions may lead to models which can produce qualitatively different and 
contradictory results.  The answer to the question `How do drops evaporate?' is still
not known.

The complete 2-sided model can be simplified by realizing that thermal conductivity and viscosity 
of vapor are small compared to the ones of the liquid.  In addition, assuming that the
gas phase is convection-free, one reduces the 2-sided model to the so-called `1.5-sided' model which includes
the processes in the liquid, and the diffusion of vapor in the surrounding air~\cite{MDD2005}.  
An estimate of a typical diffusion time scale $t_d = {l^2/D}$, involves the relevant thickness
of the gas phase, $l$, and the diffusion constant for e.g. water vapor $D \approx 10^{-5}~m^2/s$.  Assuming
for a moment that $l$ is on millimeter scale (comparable to a typical thickness
of a drop), leads to $t_d\approx 10^{-2}~s$.  The argument that $t_d$ is much shorter than
a typical time scale involved in drop evolution has been used to reduce the diffusion equation
for vapor concentration, $c$, to a Laplace equation~\cite{HL2002}.  Furthermore, assuming that
evaporation process itself is extremely fast~\cite{POPOV05} allows to completely ignore the processes
in the liquid for the purpose of finding the mass flux.  Therefore, the problem is simplified to 
$\nabla^2 c =0$ in the gas/vapor domain.  Concentrating now on the part of domain close
to the contact line, one realizes electrostatic analogy of finding an electric field
(mass flux, $J$) in the vicinity of a `lens' shaped conductor (the drop), where $c$ plays
the role of electrostatic potential~\cite{DBDHNW1997}.  Typically this lens model assumes a pinned
(stationary) contact line, although this assumption is not crucial since typically
the contact line, even if mobile, evolves slowly~\cite{CBPC2002}.  One important 
outcome of the model is that, under some additional simplifications,  $J\propto 1/h^\lambda$,
where $h$ is the drop thickness, and $\lambda=\lambda(\Theta)$, $\Theta$ being 
the contact angle~\cite{DBDHNW2000}.  Clearly, $J$ diverges at the contact line,
and various (often ad-hoc) procedures have been used to regularize the problem. 
An extensive modeling using lens-type model has been carried out, implementing both finite-element
and lubrication- type approaches~\cite{HL2002,HL2005}. Various versions of this model
have been used to predict evaporative behavior of alcane~\cite{GUENA2007}
and colloidal drops~\cite{DBDHNW2000,F2002} and the temperature along liquid-gas 
interface~\cite{bill2007}, among others.

Turning now to the liquid phase, one realizes that the evaporation is limited by two 
physical processes: heat diffusion through the liquid supplying
heat to the interface, and evaporation itself.  In a simple
model~\cite{COLINET}, these two processes can be related via Biot's number, 
$Bi = {{K p_T L d_0} / {(\rho_v k)}}$, where
$K = {\alpha \rho_v(T_i)/\sqrt{{2 \pi R_g T_{sat}}}}$ and $p_T=L p_{sat} / (R_g T_{sat}^2)$.  Here,
$p_{sat}$ is the saturation pressure, $R_g$ is the universal gas constant divided by the molar mass, 
$L$ is the latent heat of vaporization, $d_0$ is drop thickness, $\rho_v$ is 
vapor density, $k$ is liquid heat conductivity, $T_i$ and $T_{sat}$ are the interface and 
saturation temperatures, and $\alpha$ is the accommodation coefficient, describing probability of phase change.
The limit $Bi\rightarrow 0$ implies that the temperature of the liquid/gas interface
tends to the temperature of the solid, evaporation itself proceeds in reaction-limited
regime, and the interface is in a state of non-equilibrium~\cite{COLINET}.  
$Bi \rightarrow \infty$, on the other hand, indicates that the evaporation is much 
faster than the heat diffusion through the liquid, and evaporation proceeds in 
liquid heat diffusion-limited regime, with the interface being in an equilibrium state.

While most of the quantities entering the definition of $Bi$ are well known,
the value of $\alpha$ is questionable.  A variety of $\alpha$'s in
the range $O(1) - O(10^{-6})$ have been used in the literature, often without much
justification.  We note that the theoretical predictions that $\alpha \in [10^{-2},1]$ 
for water have been found to grossly overestimate the volatility, with
values in the range $[10^{-6},10^{-4}]$ being more realistic~\cite{marek2001}.
In our experiments described below, we find $\alpha \approx 10^{-4}$, consistently
with other experimental results involving drops exposed to open atmosphere~\cite{marek2001}.
This $\alpha$ gives $Bi \approx 10^{-2}$, suggesting that evaporation proceeds in the reaction-limited regime.
Further insight can be reached by considering $Bi$ as the ratio of the relevant time scales involved
in heat diffusion in the liquid, $t_l$, and the one related to the evaporative process itself, $t_e$.
Using $t_l = {d_0}^2 / \kappa$, where $\kappa$ is thermal diffusivity of liquid, and $d_0=0.5~mm$, one finds
$t_l \approx 1 sec$, giving $t_e \approx 10^2~sec$.  Similar value is obtained by using 
$t_e=(2 \rho {d_0} L)/(k \Delta T)$, where $\Delta T$ is the appropriate temperature 
scale~\cite{kondic_murisicPRL06}.
Since $t_e \gg t_l \gg t_d$, one may consider a model where relevant limiting mechanism
is the evaporative process itself, and not the diffusion of vapor in the gas.
This `non-equilibrium one sided' (NEOS) model has been extensively used for
a variety of problems involving evaporative thin films~\cite{ODB97}, but only few works have applied
it to evaporating drop problem~\cite{A2005,F2002,kondic_murisicPRL06}.
We note that the argument outlined here for use of the NEOS model is based on the assumption
of relatively small relevant thickness, $l$, governing diffusion in the gas phase.  Larger $l$ would
lead to larger $t_d$, and thus it would be unclear which model is more appropriate.  We note that, 
if $t_d$ is large, the assumption of steady-state formulation for the gas concentration is questionable.

In this letter we show computationally that these two commonly used models for evaporation 
may lead to inconsistent results regarding volume loss, time evolution of 
the drop size, and in particular regarding the temperature gradient along the liquid-gas 
interface.  We show that each model requires the use of unknown quantities, making it difficult 
to decide on which one is more appropriate.  Therefore, we carry out experiments to estimate
these quantities and compare the two models, with the goal of reaching a conclusion regarding 
an appropriate one, at least for the experiments considered.  To our knowledge, this is the first time 
these two models have been compared directly against each other using data appropriate to
a particular experiment.  

The mathematical model consists essentially of Navier-Stokes equations for the 
liquid phase coupled with the energy equation for the solid, and appropriate expression for
$J$, in the spirit of the models reviewed in~\cite{ODB97}.  
We use lubrication approximation although for some considered
problems (such as a dionized water (DIW) drop on a silicon (Si) substrate) the contact angle is 
relatively large ($\Theta\approx 40^\circ$).  This approach is supported by finite-element 
simulations that show that even for large contact angles considered here, lubrication approach leads 
to reliable results~\cite{HL2002}.   Solid-liquid interaction is included using disjoining pressure model 
with both attractive and repulsive van der Waals (vdW) terms, leading to a stable equilibrium 
thickness, which can be related to commonly used precursor film.  In cylindrical coordinates and 
assuming azimuthal symmetry, we obtain the following 4th order PDE for the film 
thickness, $h(r,t)$

\begin{eqnarray}
&&
{\partial h \over\partial t}
+ {E J}  + {S \over r} \left[r h^3 \left( h_{rrr} + {1 \over r} h_{rr} -
{1 \over r^2} h_{r} \right) \right]_r - 
\nonumber \\
&&
{E^2 \over {r D}} \left[ r J h^3 J_r \right]_r
+ {M \over {r P}} \left[ r h^2 \left( h + \mathcal{W} \right) J_r \right]_r
+ {M \over {r P}} \left[ r J h^2 h_r \right]_r
\nonumber
\\ 
&&
+ {A \over r} \left[ r h^3 \left( \left({b \over h}\right)^3 -
\left({b \over h}\right)^2\right)_r\right]_r
+ {G \over r} \left[ r h^3  h_r \right]_r = 0\, ,
\label{eq:h}
\end{eqnarray}
which is put in nondimensional form by choosing $d_0$ and $d_0^2/\nu$ as the length and 
time scale, respectively ($\nu$ is the kinematic viscosity).
We used a similar formulation to consider instabilities of evaporative isopropyl alcohol (IPA) 
drops~\cite{kondic_murisicPRL06}.  The main difference of the present case
(in addition to the geometry) is keeping the evaporative flux $J(h)$ explicitly
in the formulation, so that Eq.~(\ref{eq:h}) can be used for any evaporative model.
Eq.~(\ref{eq:h}) includes the effects due to viscosity, evaporation, capillarity, vapor recoil, Marangoni and 
vdW forces, and gravity. 
All nondimensional quantities are defined in~\cite{kondic_murisicPRL06}, and here we just
enumerate the most important ones: $E, S, M, P, A, G$ correspond to evaporation number, 
nondimensional surface tension, Marangoni number, Prandtl number, nondimensional Hamaker constant, 
and gravity.  The material parameters are listed in~\cite{kondic_murisicPRL06}.

Next we concentrate on the key point, and that is the evaporative flux, $J$.  For the lens model, we
use $J(h) = {\chi / {h^\lambda}}$, where the exponent $\lambda$ can be approximated 
by $\lambda =  0.5 - {\Theta/\pi}$~\cite{HL2002}.  For the NEOS model, we use 
$J(h) = {1 / {(h+\mathcal{W}+\mathcal{K})}}$~\cite{kondic_murisicPRL06},
where $\mathcal{W}$ accounts for finite thermal conductivity 
of the solid substrate and $\mathcal{K} = Bi^{-1}$.  For the case of DIW, $\mathcal{K}$ is
typically large, as discussed before, and therefore $J(h)$ only weakly depends on $h$.
The crucial parameters, volatility $\chi$ and $\alpha$ are obtained from the experiments 
described next.  

\begin{figure}[thb]
\centering
\subfigure{\includegraphics[height=25mm,width=40mm]{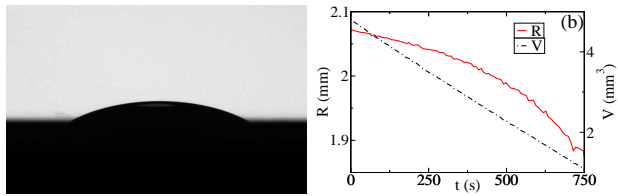}}
\subfigure{\includegraphics[height=25mm,width=40mm]{figure1b.eps}}
\caption{
(a) Snapshot of evaporating DIW drop on Si substrate; (b) Volume and
radius of the drop as a function of time (experimental results).
}
\label{fig:exp}
\end{figure}

The experiments are carried out at room temperature and in open atmosphere, 
using a goniometer (KSV CAM $200$) which consists of a camera,
light source, and static deposition platform.
Pure liquids and silicon wafers of semiconductor grade smoothness are used.
Figure~\ref{fig:exp}a) shows an example of an image of evaporating DIW drop.
The build-in software analyzes the images and yields the height and the radius of a drop.  
Figure~\ref{fig:exp}b) shows the resulting radius and volume of evaporating drop as a 
function of time.   In agreement with other works, we find linear decrease of the
volume for the considered time interval~\cite{DBDHNW2000,D2000,HL2002}.
We note that we do not observe either significant stick-slip
motion, nor contact line pinning that is often assumed to occur~\cite{HL2002,HL2005}.  Possibly, 
these effects are not seen due to small surface roughness of the Si wafers and
due to purity of the liquids.

\begin{figure}[thb]
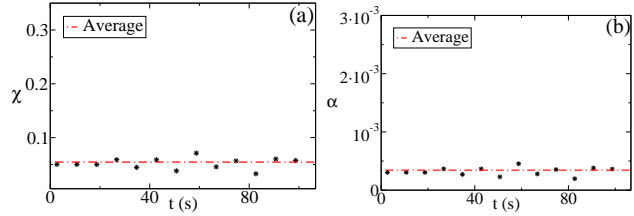

\centering
\epsfxsize=1.6in
\subfigure
{\epsffile{figure2a.eps}}
\epsfxsize=1.6in
{\epsffile{figure2b.eps}}
\caption{
Volatility for (a) lens and (b) NEOS models for DIW/Si, leading to 
$\chi \approx 5.5 \cdot 10^{-2}$ and $\alpha \approx 3.4 \cdot 10^{-4}$ (lines).
}
\label{fig:volatility}
\end{figure}

In order to extract $\chi$ and $\alpha$ from the experimental data, 
we proceed as follows.  The considered experimental time frame (typically the first $106~sec$), 
is split into subintervals, and then the mass loss in each subinterval is used to obtain 
appropriate values of $\chi$ and $\alpha$ by integrating $J$ over the drop surface (more
details regarding this procedure will be given elsewhere~\cite{mk_future}).  
We expect that this method is more accurate than
simply using the dry out time to estimate the flux, since at late stages of evolution, 
the evaporation rate may be reduced~\cite{D2000}.  Figure~\ref{fig:volatility} shows the 
results.  For the purpose of verification, we have applied the same approach to
the already available experimental data~\cite{HL2002} and obtained similar values for $\alpha$ and $\chi$.
With these values available, we are now almost ready to proceed with 
numerical simulations of the drop evolution governed by Eq.~(\ref{eq:h}).  The only missing 
ingredient is the precursor film thickness, $b$.  Although it is typically sufficient to
use $b\ll 1$, here we are more careful since quantitative agreement is desired.  We are 
governed by the requirement that $r(t)$ and $V(t)$ should not depend on $b$ in any significant 
manner, and have found that for $d_0b \le 0.625~\mu m$ this requirement is 
satisfied.  Coincidentally or not, this value is consistent with the equilibrium adsorbed film thickness 
$d_0 b_e$ for which evaporation stops due to attracting vdW forces~\cite{A2005}.  
For the appropriate parameters, we find $d_0b_e\approx 0.5~\mu m$. 
We note that consideration of vdW forces automatically 
regularizes otherwise singular expression for $J$ in the lens model~\cite{saritha_pof07}.  
Without inclusion of vdW effects, an additional externally added regularization of $J$ has to 
be included in order to correctly compute the total mass flux.  

\begin{figure}[thb]
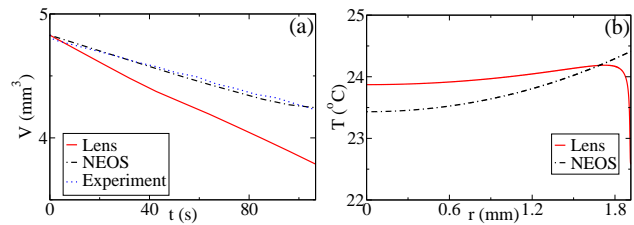

\centering
\epsfxsize=1.6in
\subfigure
{\epsffile{figure3a.eps}}
\epsfxsize=1.6in
{\epsffile{figure3b.eps}}
\caption{(a) Volume of an evaporating DIW drop: comparison of the models with the experiment. (b) Temperature
of the liquid-gas interface predicted by the two models at the final time shown in (a).
}
\label{fig:vol_temp}
\end{figure}

The numerical simulation of Eq.~(\ref{eq:h}) is carried out using second order accurate 
implicit scheme which is an extension of the one used in~\cite{kondic_murisicPRL06} to 
cylindrical geometry.  All simulations use as an initial condition the steady-state solution
of Eq.~(\ref{eq:h}) obtained by removing the evaporative terms; in experiments, this configuration
is reached after a very short time.
Figure~\ref{fig:vol_temp}a) shows the volume of the evaporating drop as a function of time. 
We find very good agreement between the experimental results and the NEOS model, 
while the lens model overestimates the volume loss due 
to evaporation.  Therefore we conclude that, at least for the problem considered, the NEOS 
model predicts better the volume loss compared to the lens model.   We discuss this difference,
as well as additional experiments/simulations involving evaporation on different
solid surfaces and under modified experimental conditions elsewhere~\cite{mk_future}.

Figure~\ref{fig:vol_temp}b) shows that the two models predict qualitatively different temperature profiles 
along the liquid-gas interface.  An increase of temperature as one moves from the center 
in the NEOS model is the consequence of the fact that the heat supplied from the solid is larger than the heat 
loss due to evaporation.
The lens model, on the other hand, predicts significantly larger evaporative flux in the 
contact line region, therefore leading to a sharp decrease of temperature.  An increase of 
temperature as one moves away from the drop center is consistent with the results~\cite{bill2007,HL2005b} obtained 
using the lens model, and similar values of $\Theta$ and thermal conductivities of the liquid and solid, although
under the assumption of a pinned contact line.  However, the results presented here predict a `stagnation point', 
where the temperature gradient changes sign.  Presence of a stagnation point, although based on different 
physical grounds, was discussed earlier as one of the necessary ingredients required for formation of ring-like 
deposits occurring during evaporation of colloidal drops~\cite{DBDHNW2000}.   We note that 
simulations assuming the lens model for smaller values of $\Theta$ lead to monotonically decreasing temperatures 
along the liquid-gas interface as one moves away from the drop center, in full agreement with the earlier 
results~\cite{bill2007,HL2005b}. 

Next, we apply these two models to a more volatile IPA drop. 
This configuration is characterized by $\Theta \approx 6^\circ$ leading to very thin
drops.  Due to this fact, in our experiments we cannot follow accurately the evolution in the manner
it was done for DIW drops.  Therefore, we use the dryout time, which we 
can estimate with a reasonable accuracy, to obtain the required values of $\alpha$ and $\chi$.  
For $V(0) \approx 3.2~mm^3$, we find $t_{dry} \approx 135~sec$.  Applying the two evaporative models
leads to $\alpha = 9.3 \cdot 10^{-4}$ and $\chi = 5.7 \cdot 10^{-3}$.  We note that 
similar values of $\chi$ were obtained in experiments with pure alcanes~\cite{GUENA2007}.
Implementing these values, we solve Eq.~(\ref{eq:h}) using the parameters appropriate for 
IPA and Si~\cite{kondic_murisicPRL06}.  

\begin{figure}[thb]
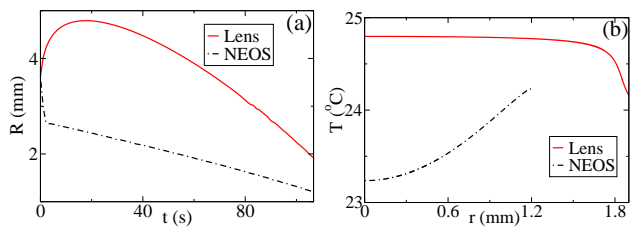

\centering
\epsfxsize=1.6in
\subfigure
{\epsffile{figure4a.eps}}
\epsfxsize=1.6in
{\epsffile{figure4b.eps}}
\caption{IPA drop: (a) Radius $R(t)$ as a result of the two models; (b) 
Temperature profiles of the liquid-gas interface at the final time shown in (a).
}
\label{fig:ipa}
\end{figure}

Figure~\ref{fig:ipa}a) show the resulting evolution of the drop radius, $R(t)$, and the temperature 
along the liquid-gas interface (b).  First, we notice dramatically different evolution of $R(t)$ for the 
two models.  Considering the temperature profiles in~\ref{fig:ipa}b) provides immediate 
understanding of this difference.  The Marangoni forces act in the opposing directions,
leading to a very different evolution.  For example, for the lens model, the Marangoni
forces act outwards, leading to initial increase of the drop radius (shown in Fig.~\ref{fig:ipa}a)  
despite the loss of mass due to evaporation.  The difference between the temperature profiles 
predicted by the two models is much more significant in the case of IPA compared to DIW due to its 
larger volatility.  We discuss the details of model predictions regarding evolution of IPA drops in 
more details elsewhere~\cite{mk_future}.

We find it intriguing that the two commonly used models for evaporation produce results
that are qualitatively different, regarding the evolution of drops' volume, 
radius, and, in particular, the liquid-gas interfacial temperature.  The 
resulting Marangoni forces may act in the opposing directions.  Therefore, it appears 
that the arguments involving, e.g., the influence of Marangoni forces on formation of 
particle deposits next to the contact line have to be carefully reconsidered.  We hope 
that the results presented here will encourage more elaborate experiments, possibly 
involving direct measurement of the interfacial liquid-gas temperature, an 
ultimate test for any evaporation model.

{\it Acknowledgments.} We thank Pierre Colinet, Javier Diez, Yehiel Gotkis, Alex Oron, 
Bill Ristenpart, and Howard Stone for useful discussions.


\end{document}